\documentclass[aps,prl,twocolumn,showpacs,groupedaddress,floatfix]{revtex4}

\usepackage{graphicx}
\usepackage{dcolumn}
\usepackage{epsfig}
\usepackage{amssymb}
\usepackage{amsmath}
\usepackage{color}

\begin{document}

\title{Signature of elasticity in the Faraday instability}

\author{Pierre Ballesta\footnote{Corresponding author: {\tt ballesta@crpp-bordeaux.cnrs.fr}}}
\affiliation{Centre de Recherche Paul Pascal, Avenue Schweitzer, 33600 Pessac, FRANCE}
\author{S\'ebastien Manneville}
\affiliation{Centre de Recherche Paul Pascal, Avenue Schweitzer, 33600 Pessac, FRANCE}

\date{\today}
\begin{abstract}
We investigate the onset of the Faraday instability in a vertically vibrated wormlike micelle solution.
In this strongly viscoelastic fluid,
the critical acceleration and wavenumber are shown to present oscillations as a function of driving frequency
and fluid height. This effect, unseen neither in simple fluids nor in previous experiments on polymeric fluids, 
is interpreted in terms of standing elastic waves between the disturbed surface and the container bottom.
It is shown that the model of S. Kumar [Phys. Rev. E, {\bf 65}, 026305 (2002)] for a viscoelastic
fluid accounts qualitatively for our experimental observations.
Explanations for quantitative discrepancies are proposed,
such as the influence of the nonlinear rheological behaviour
of this complex fluid.
\end{abstract}
\maketitle

Since Faraday's founding work \cite{Faraday:1831}, the parametric instability of a vertically vibrated
fluid layer has emerged as one of the best candidates to study pattern formation and nonlinear dynamics.
Above a critical acceleration $a_c$, an initially flat and quiescent fluid layer driven vertically 
at frequency $f$ goes unstable and gives way to a pattern of surface waves that oscillates at half
the driving frequency with a characteristic wavenumber $k_c$
\cite{Faraday:1831,Benjamin:1954,Chen:1997}. As the driving acceleration
is raised above onset, a series of secondary instabilities take place, leading to defect dynamics and eventually
to spatio-temporal chaos \cite{Kudrolli:1996}.

So far, the Faraday instability has been mostly studied in {\it simple} viscous fluids
\cite{Faraday:1831,Benjamin:1954,Chen:1997,Kudrolli:1996,Edwards:1994,Kumar:1994}. 
Recently, interest has grown in the effect of vertical vibrations on a layer of {\it complex} fluid,
both experimentally \cite{Raynal:1999,Wagner:1999} and theoretically \cite{Kumar:1999,Muller:1999}.
Due to their microstructure, complex fluids display viscoelastic properties,
which may affect classical hydrodynamic instabilities \cite{Larson:1999}.
However, previous experiments on semidilute polymeric solutions did not show significant modifications of the
Faraday instability \cite{Raynal:1999,Wagner:1999}. The debate has mainly focused on the existence
of a harmonic response (at $f$) instead of the classical subharmonic response (at $f/2$), a typical viscoelastic 
effect predicted numerically \cite{Kumar:1999,Muller:1999} and observed experimentally
at rather low frequencies ($f<40$~Hz) together with new types of patterns that compete with
each other \cite{Wagner:1999}. 

\begin{figure}[h]
\begin{center}
\scalebox{0.9}{\includegraphics{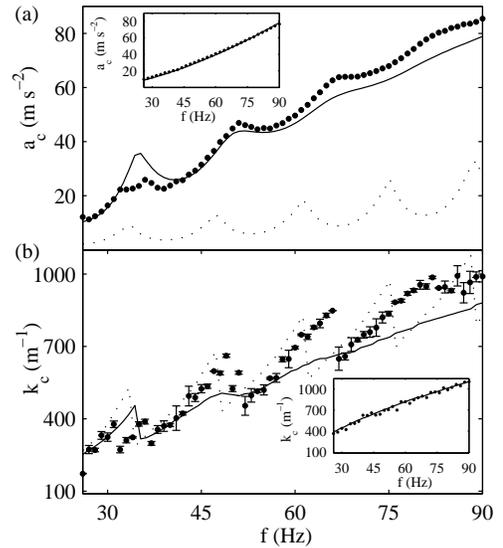}}
\end{center}
\caption{(a) Critical acceleration $a_c$ and 
(b) critical wavenumber $k_c$ vs driving frequency in a 4~\% wt. CPCl/NaSal wormlike micelle solution
($\bullet$). Also shown are the calculations for the corresponding Maxwell fluid (dotted lines) 
and when adding a Zimm-like term (solid lines).
Insets: $a_c$ and $k_c$ for a Newtonian fluid (78~\% glycerol--22~\% water mixture)
and corresponding predictions using the model of Ref.~\cite{Kumar:1994} with density
$\rho=1.19$~g\,cm$^{-3}$, viscosity $\eta=0.05$~Pa\,s, and surface tension $\sigma=0.06$~N\,m$^{-1}$
(solid lines).}
\label{f.omexp}
\end{figure}

In this Letter, we report onset measurements of the Faraday instability
in a wormlike micelle solution, a strongly
viscoelastic fluid well characterized by a single relaxation time $\tau$.
The critical acceleration and wavenumber are shown to present {\it oscillations} as a function of driving frequency (see Fig.~\ref{f.omexp}). This effect is interpreted in terms of 
standing elastic waves between the disturbed surface and the container bottom.
It is also shown that
the finite-depth model of Ref.~\cite{Kumar:1999} accounts qualitatively for these experimental observations.
Finally, possible explanations for discrepancies are discussed and the implications of our results in the
field of complex fluid hydrodynamics are emphasized.

Wormlike micelles (also called ``living polymers'') are long, cylindrical aggregates of
surfactant molecules in solution \cite{Larson:1999}.
They spontaneously form under given conditions of temperature and concentration for a wide range of
surfactants. Unlike conventional polymers, their size is not
fixed and they constantly break and recombine under thermal agitation. Their dynamics thus result
from both a classical reptation motion \cite{Bird:1987} and breaking/recombination processes \cite{Cates:1987}. This
peculiar feature leads to an almost perfect Maxwellian behaviour in the small-deformation regime,
for which the complex viscosity reads: $\eta_M(\omega)=G_0\tau/(1+i\omega\tau)$,
where $G_0$ is the shear modulus, $\tau$ the relaxation time,
and $\omega$ the pulsation.

\begin{figure}[htbp]
\begin{center}
\scalebox{0.9}{\includegraphics{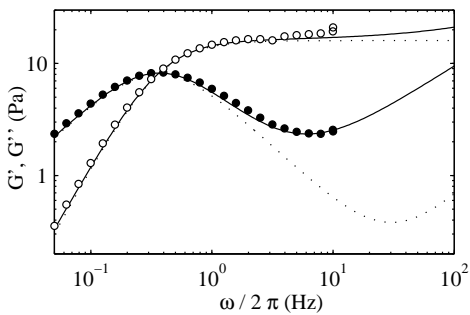}}
\end{center}
\caption{Storage modulus $G'(\omega)$ ($\circ$) and loss modulus $G''(\omega)$ ($\bullet$)
as a function of oscillation frequency $\omega/2\pi$ for a stress amplitude of 2~Pa. The dotted line is the best
Maxwell fit ($G_0=16$~Pa and $\tau=0.44$~s) with a solvent viscosity $\eta_s=10^{-3}$~Pa\,s. 
The solid line corresponds to the same Maxwell model with an added Zimm-like term (see text).}
\label{f.rheo}
\end{figure}

Our working fluid is a wormlike micelle system made of cetylpyridinium chloride (CPCl, from Aldrich) and
sodium salicylate (NaSal, from Acros Organics) dissolved in brine (0.5~M NaCl) \cite{Rehage:1988,Berret:1997}. In this study, we focus on a 4~\% wt. sample
with a concentration ratio [NaSal]/[CPCl]=0.5 as described in Ref.~\cite{Berret:1997}. Linear rheological
measurements were performed in the cone-and-plate geometry using a stress-controlled rheometer (AR~2000N, TA Instruments) with small stress
oscillations of amplitude 2~Pa (the linear regime extends up to 20~Pa). As shown
in Fig.~\ref{f.rheo}, the dynamic moduli are well accounted for by a Maxwell model
$G'(\omega)=-\omega\Im(\eta_M(\omega))$ and
$G''(\omega)=\omega\eta_s+\omega\Re(\eta_M(\omega))$, with
$G_0=16$~Pa and $\tau=0.44$~s, at least for frequencies below 2~Hz.
$\eta_s=10^{-3}$~Pa\,s represents the solvent (brine) contribution
to the viscosity.

Our experimental setup is a classical one for studying the Faraday instability. 
A cylindrical container of inner diameter 60~mm is filled to a height $h=10$~mm 
under the brimful boundary condition \cite{Edwards:1994} and vertically vibrated using an electromagnetic shaker
(Ling Dynamic Systems V406). 
The fluid temperature is regulated to $21\pm 0.05^{\circ}$C by water circulation beneath the container.
In order to prevent evaporation and surface contamination,
the container is sealed by a Plexiglas cover that supports a small piezoelectric
accelerometer (Endevco 2224C). The signal from the accelerometer is fed to a lock-in amplifier
(Stanford Research Systems SR810) that provides the sinusoidal excitation at frequency $f$ to the shaker. 
The experiment is illuminated from above by a circle of diodes strobed at $f$ or $f/2$.
The bottom of the container is made of aluminum to provide mechanical rigidity,
and anodized to prevent light reflection.

For a given driving frequency, the instability threshold
is determined by (i) noting the acceleration for which the instability
first appears, (ii) fully destabilizing the whole surface by increasing the acceleration about 5~\% above onset,
and (iii) decreasing the acceleration and noting the acceleration for which the instability completely disappears.
In all cases, the two values of the critical acceleration differ by less than 2~\%: no significant hysteresis is observed.
The value of $a_c$ shown in Fig.~\ref{f.omexp}(a) is the mean of these two accelerations, their difference remaining always
smaller than the marker size. Moreover, the surface response was found to be {\it subharmonic} (at $f/2$) over the
whole range of investigated frequencies $f=25$--90~Hz.
For each frequency, the critical wavenumber $k_c$ is estimated from a picture of the fully destabilized surface.

Figure~\ref{f.omexp} constitutes our main result. 
The critical acceleration and wavenumber are not monotonic but
rather oscillate with the frequency, a maximum in $a_c$ corresponding to a large drop of $k_c$.
To our knowledge, such a marked effect has not been reported in previous experiments
on complex fluids. In Newtonian fluids, non-monotonic $a_c$ or $k_c$ are associated
to {\it lateral} boundary effects (as only an integer or semi-integer number of wavelengths may
fit in the cell when the brimful condition is used) \cite{Edwards:1994,Ciliberto:1984}.
In order to rule out such an interpretation,
we repeated the experiment with a Newtonian fluid
of critical acceleration and wavenumber similar to our wormlike micelle solution. 
The insets in Fig.~\ref{f.omexp} show monotonic behaviours, in perfect agreement with
the model of Ref.~\cite{Kumar:1994} for a laterally unbounded viscous fluid,
as already found in other Newtonian fluids \cite{Edwards:1994}. This allows us
to consider our results in terms of a laterally unbounded fluid and
confirms the viscoelastic nature of the oscillations seen in $a_c$ and $k_c$ for the
wormlike micelles.

More precisely, we propose to interpret these oscillations as a coupling between the disturbed surface
and elastic waves reflected at the container bottom. For a Maxwell fluid in the $\omega\tau\gg 1$ limit, which is always verified in our experiments,
the velocity $c$ and attenuation coefficient $\alpha$ of shear
waves are given by
$c=\sqrt{G_0/\rho}$ and $\alpha=1/(2 c \tau)$, 
where $\rho$ is the fluid density. Above onset, surface disturbances generate elastic waves
that may form a standing wave in the container height $h$. When $2h$ is a semi-integer
multiple of the wavelength $2c/f$ \cite{Remark}, i.e.
$f=(n+1/2)f_1$ with $f_1=\sqrt{G_0/\rho}/h$ and $n$ an integer, constructive interferences form
at the surface. In this case, elastic waves are amplified and surface disturbances are promoted
so that the critical acceleration is expected to be lower than in the absence of elastic effects.
On the other hand, when $f=nf_1$, destructive interferences should lead to an increase of $a_c$,
hence the oscillations of Fig.~\ref{f.omexp}(a). Moreover, at each maximum of $a_c$, the number
of elastic modes in the vertical direction increases. Since the vertical wavenumber is linked to
the horizontal one through the incompressibility condition \cite{Kumar:1994,Kumar:1999}, jumps similar to those
of Fig.~\ref{f.omexp}(b) are expected even in a laterally unbounded fluid. 

\begin{figure}[htbp]
\begin{center}
\scalebox{1.1}{\includegraphics{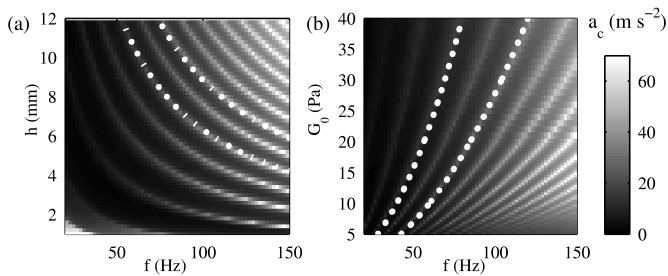}}
\end{center}
\caption{Numerical calculations using the
model of Ref.~\cite{Kumar:1999} with $\tau=0.44$~s,
$\rho=1.0$~g\,cm$^{-3}$, and $\sigma=0.03$~N\,m$^{-1}$. (a) $a_c$ vs $h$ and $f$
at fixed $G_0=16$~Pa. (b) $a_c$ vs $G_0$ and $f$ at fixed $h=10$~mm.
White dots are the lines $f_n=n\sqrt{G_0/\rho}/h$ with $n=5$ and 7 and coincide with maxima of $a_c$.}
\label{f.numcomp}
\end{figure}

In order to further check the above interpretation, we performed the finite-depth numerical calculations 
of Ref.~\cite{Kumar:1999} for a Maxwell fluid 
with the rheological parameters inferred from Fig.~\ref{f.rheo}. The two
remaining parameters used in the calculations, namely the density and the surface tension of the fluid,
were measured independently yielding $\rho=1.0$~g\,cm$^{-3}$ and $\sigma=0.03$~N\,m$^{-1}$ respectively.
By varying $h$ ($G_0$ resp.) for a fixed $G_0$ ($h$ resp.), we show in Fig.~\ref{f.numcomp} that
large oscillations also show up numerically. Moreover, they perfectly
agree with the simple relation $f_n=n\sqrt{G_0/\rho}/h$ suggested above.
The interpretation in terms of elastic waves reflected at the container bottom is thus confirmed.
This {\it finite-depth} viscoelastic effect went unseen in previous numerical studies that
focused on the question of harmonic vs subharmonic response \cite{Kumar:1999,Muller:1999}.

Note, however, that these numerical calculations based on the Maxwell model
do not allow any quantitative comparison with the experimental $a_c$ and $k_c$
(see dotted lines in Fig.~\ref{f.omexp}). This is most probably because our fluid does not remain purely Maxwellian
at frequencies higher than 2~Hz. Indeed, although our rheological measurements are limited to 10~Hz,
an upturn of $G''(\omega)$ is clearly visible on the data of Fig.~\ref{f.rheo}. Such an upturn
is usually attributed to the presence of Rouse-like or Zimm-like motion 
at high frequencies \cite{Bird:1987,Fischer:1997}.
If we model this behaviour by adding a Zimm relaxation term to the previous Maxwell
model,
\begin{equation}
\eta(\omega)=\eta_s+\frac{G_0 \tau}{1+i\omega\tau}+a_Z\left(1-\frac{i}{\sqrt{3}}\right)\omega^{-\frac{1}{3}}\,,
\label{e.zimm}
\end{equation}
a very good fit of both $G'(\omega)$ and $G''(\omega)$ is obtained with $a_Z=0.12$~Pa\,s$^{2/3}$ (see solid lines in Fig.~\ref{f.rheo}).

Using Eq.~(\ref{e.zimm}) in the numerical calculations leads to computed curves
(shown as solid lines in Fig.~\ref{f.omexp}) that are much closer to the experimental
measurements, although the attenuation of the oscillations with increasing frequency is now too strong (see discussion below).

\begin{figure}[htbp]
\begin{center}
\scalebox{0.9}{\includegraphics{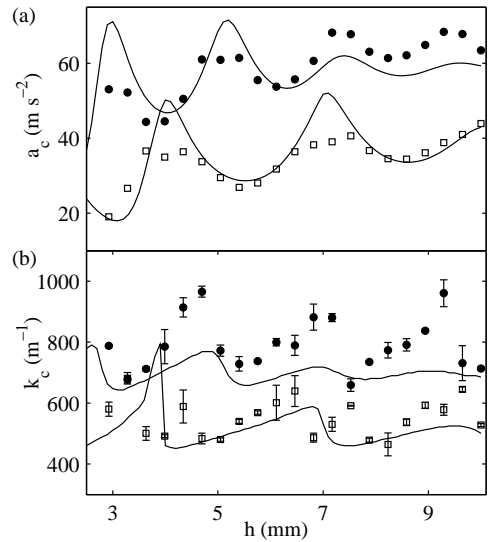}}
\end{center}
\caption{(a) Critical acceleration $a_c$ and (b) critical wavenumber $k_c$ vs fluid height
measured at 50~Hz ($\square$) and at 70~Hz ($\bullet$). The solid lines correspond to the Maxwell
model with an added Zimm-like term (see text).}
\label{f.hexp}
\end{figure}

Finally, both our simple analysis and the numerical results of Fig.~\ref{f.numcomp}(a) predict
that $a_c$ and $k_c$ should oscillate when the fluid height $h$ is varied at a fixed frequency. Figure~\ref{f.hexp} shows that such
oscillations indeed take place in the experiment and that the Maxwell model corrected by a Zimm-like term
provides a good description of $a_c$. Again, these findings are strikingly different from the monotonic decay of $a_c$ vs $h$ in Newtonian fluids \cite{Edwards:1994}.

Let us now discuss the present experimental and numerical results.
Previous experiments on polymeric solutions did not show any oscillations in the onset
measurements \cite{Raynal:1999,Wagner:1999}. This can be explained by considering the attenuation
coefficient of the elastic waves $\alpha$. If $\alpha> 1/h$, no standing wave, and
thus no oscillation, is observed due to a too strong attenuation. In our wormlike micelle solution,
the storage modulus $G'(\omega)$ remains always at least three times larger than the loss
modulus $G''(\omega)$ and
$\alpha h\simeq 0.09$, so that oscillations are observed, whereas 
previous experiments were performed either in the capillary regime \cite{Raynal:1999}
or in the overdamped regime \cite{Wagner:1999}
where elastic waves may not be excited \cite{Harden:1991}.

Moreover, even though the oscillations in $a_c$ are rather well reproduced by the Maxwell model with a
Zimm correction, the agreement with the experimental values of $k_c$ remains only qualitative (see Figs.~\ref{f.omexp}(b)
and \ref{f.hexp}(b)). A Rouse correction was also tried with similar results.
It may be argued that the discrepancy is due to an inadequate rheological model. Indeed,
standard rheometers are limited to about 15~Hz,
so that rheological data had to be extrapolated to higher frequencies \cite{Raynal:1999,Wagner:1999}.

However, we believe that the lack of agreement points to more fundamental questions about the
validity of the numerical approach. Indeed, besides lateral finite-size effects, which we ruled out at
the beginning of this study, at least two important physical effects are not taken into account
in the model of Ref.~\cite{Kumar:1999}: interfacial rheology and nonlinear bulk rheology.

First, in the numerical calculations, the air--fluid interface is modelled as purely elastic with surface tension
$\sigma$, whereas in a complex fluid, $\sigma$ can depend on the frequency. More importantly,
the {\it interfacial viscoelasticity} may significantly affect the boundary conditions \cite{Edwards:1991}.

Second, the model of Ref.~\cite{Kumar:1999} relies on linear viscoelasticity. Indeed,
since the instability occurs from a quiescent state, strain is expected to be
``small.'' Experimentally, the smallest detectable surface deformation is $\zeta\simeq 10~\mu$m.
The corresponding strain rate $\dot{\gamma}$ may be estimated by $\dot{\gamma}\sim\zeta\omega k\simeq 10$~s$^{-1}$ at $f=60$~Hz \cite{Kumar:1999}. Since the zero-shear viscosity of our fluid
is $\eta_0=G_0\tau\simeq 7$~Pa\,s, $\dot{\gamma}\simeq 10$~s$^{-1}$ corresponds to a shear stress of about 70~Pa, which is far into the 
{\it nonlinear regime}. Thus, the only knowledge of the linear viscoelastic moduli $G'$ and $G''$ may
not be sufficient to fully account for the experiments and nonlinear effects could provide an
explanation for the observed discrepancies.

More precisely, wormlike micelles are known to align under shear, leading to a shear-induced nematic
state \cite{Larson:1999}. This strongly {\it shear-thinning} isotropic-to-nematic transition occurs
for $\dot{\gamma}\simeq 0.5$--50~s$^{-1}$ depending on temperature and concentration \cite{Berret:1997}.
In a Faraday wave pattern, shear is localized
between the crests and the troughs. Thus, above onset, aligned micelles are expected to coexist with the isotropic state
and to spatially mimic the surface pattern.
Finally, since the destabilized state is far from a pure shear flow, one may also wonder
about the influence of nonlinear {\it extensional} rheology.

To conclude, the present study reveals a strong signature of elasticity on the onset of the Faraday instability
in a wormlike micelle solution. This original effect results from standing elastic waves in the container height and should be very general in highly viscoelastic fluids.
It shows that, as suggested earlier \cite{Wagner:1999,Kumar:1999}, the characteristic relaxation time
of a complex fluid may couple to the excitation period and/or to memory effects,
leading to new temporal and spatial behaviours under vertical vibrations.
Linear viscoelasticity allows for a good qualitative agreement with the model of
Ref.~\cite{Kumar:1999} but we believe that surface rheology and/or nonlinear effects are also significant.
This raises new challenges for theory and modelling of hydrodynamic instabilities in complex fluids.
In particular, we point out that the presence of a shear-induced isotropic-to-nematic transition could play a major role
in the pattern selection. This last remark has to be related to some very
recent experimental findings on vertically vibrated shear-thickening suspensions, where a rich variety
of patterns such as holes and fingers was observed after a finite perturbation was applied to the surface
\cite{Merkt:2004}. Such an unusual behaviour was linked to the nonlinear rheological properties of the fluid.
Our results on wormlike micelles provide another
example of a striking effect induced by the microstructure of a complex fluid
on a classical instability.

\begin{acknowledgments}
The authors wish to thank A.~Colin, F.~Molino, G.~Ovarlez, and R.~Wunenburger for fruitful discussions
and the ``Cellule Instrumentation'' of CRPP for technical advice and design of the experiment. 
\end{acknowledgments}


\begin{thebibliography}{40}

\bibitem{Faraday:1831}
M.~Faraday,
Philos. Trans. R. Soc. Lond., {\bf 52}, 319 (1831).

\bibitem{Benjamin:1954}
T.~B.~Benjamin and F.~Ursell,
Proc. R. Soc. London A, {\bf 225}, 505 (1954).

\bibitem{Chen:1997}
P.~Chen and H.~Vi{\~n}als,
Phys. Rev. Lett., {\bf 79}, 2670 (1997).

\bibitem{Kudrolli:1996}
A.~Kudrolli and J.~P.~Gollub,
Physica D, {\bf 97}, 133 (1996).

\bibitem{Edwards:1994}
W.~S.~Edwards and S.~Fauve,
J. Fluid Mech., {\bf 278}, 123 (1994);
J.~Bechhoefer, V.~Ego, S.~Manneville, and B.~Johnson,
J. Fluid Mech., {\bf 288}, 325 (1994);
E.~A.~Cerda and C.~T.~Tirapegui,
J. Fluid Mech., {\bf 368}, 195 (1998).

\bibitem{Kumar:1994}
K.~Kumar and L.~S.~Tuckerman,
J. Fluid Mech., {\bf 279}, 49 (1994).

\bibitem{Raynal:1999}
F.~Raynal, S.~Kumar, and S.~Fauve,
Eur. Phys. J. B, {\bf 9}, 175 (1999).

\bibitem{Wagner:1999}
C.~Wagner, H.~W.~M{\"u}ller, and K.~Knorr,
Phys. Rev. Lett., {\bf 83}, 308 (1999).

\bibitem{Kumar:1999}
S.~Kumar,
Phys. Fluids, {\bf 11}, 1970 (1999);
Phys. Rev. E, {\bf 65}, 026305 (2002).

\bibitem{Muller:1999}
H.~W.~M{\"u}ller and W.~Zimmermann,
Europhys. Lett., {\bf 45}, 169 (1999).

\bibitem{Larson:1999}
R.~G. Larson, {\em The Structure and Rheology of Complex Fluids}
(Oxford University Press, Oxford, 1999).

\bibitem{Bird:1987}
R.~B.~Bird, R.~C.~Armstrong, and O.~Hassager, 
{\em Dynamics of Polymeric Liquids}
(Cambridge University Press, Cambridge, 1987).

\bibitem{Cates:1987}
M.~E.~Cates,
Macromolecules, {\bf 20}, 2289 (1987).

\bibitem{Rehage:1988}
H.~Rehage and H.~Hoffmann,
J. Phys. Chem., {\bf 92}, 4712 (1988).

\bibitem{Berret:1997}
J.~F.~Berret, G.~Porte, and J.~P.~Decruppe,
Phys. Rev. E, {\bf 55}, 1668 (1997).

\bibitem{Ciliberto:1984}
S.~Ciliberto and J.~P.~Gollub,
Phys. Rev. Lett., {\bf 52}, 922 (1984).

\bibitem{Remark}
The factor 2 in $2h$ accounts for the reflection on the container bottom while the factor
2 in $2c/f$ corresponds to the wavelength of subharmonic elastic waves (at $f/2$).

\bibitem{Fischer:1997}
P.~Fischer and H.~Rehage,
Langmuir, {\bf 13}, 7012 (1997).

\bibitem{Harden:1991}
J.~L.~Harden, H.~Pleiner, and P. A. Pincus,
J. Chem. Phys., {\bf 94}, 5208 (1991).

\bibitem{Edwards:1991}
D.~A.~Edwards, H.~Brenner, D.~T.~Wasan,
{\em Interfacial Transport Processes and Rheology}
(Butterworth-Heinemann, Stonheam, 1991).

\bibitem{Merkt:2004}
F.~S.~Merkt, R.~D.~Deegan, D.~I.~Goldman, E.~C.~Rericha, and H.~L.~Swinney,
Phys. Rev. Lett., {\bf 92}, 184501 (2004).

\end{thebibliography}
\end{document}